\documentclass[conference]{IEEEtran}
\IEEEoverridecommandlockouts
\usepackage{cite}
\usepackage{amsmath,amssymb,amsfonts}
\usepackage{caption}
\usepackage{algorithm}
\usepackage{algpseudocode}
\usepackage{graphicx}
\usepackage{textcomp}
\usepackage{xcolor}
\usepackage{enumitem}

\algrenewcommand\algorithmicindent{0.5em}
\def\BibTeX{{\rm B\kern-.05em{\sc i\kern-.025em b}\kern-.08em
    T\kern-.1667em\lower.7ex\hbox{E}\kern-.125emX}}
\begin{document}

    \makeatletter
    \newcommand{\linebreakand}{%
      \end{@IEEEauthorhalign}
      \hfill\mbox{}\par
      \mbox{}\hfill\begin{@IEEEauthorhalign}
    }
    \makeatother

\title{Dynamic Power and Frequency Allocation Scheme for Autonomous Platooning
\thanks{The work has been realized within the project no. 2018/29/B/ST7/01241 funded by the National Science Centre in Poland.}
}

\author{\IEEEauthorblockN{Pawe\l~Sroka}
\IEEEauthorblockA{\textit{\small Poznan University of Technology}\\
\small Poznan, Poland \\
\small pawel.sroka@put.poznan.pl}
\and
\IEEEauthorblockN{Adrian~Kliks}
\hspace{-1.5cm}\IEEEauthorblockA{\textit{\small Poznan University of Technology}\\
\small Poznan, Poland \\
\small adrian.kliks@put.poznan.pl}
\and
\IEEEauthorblockN{Micha{l}~Sybis}
\hspace{-1.5cm}\IEEEauthorblockA{\textit{\small Poznan University of Technology}\\
\small Poznan, Poland \\
\small michal.sybis@put.poznan.pl}
\and
\IEEEauthorblockN{Pawe\l~Kryszkiewicz}
\hspace{-1.5cm}\IEEEauthorblockA{\textit{\small Poznan University of Technology}\\
\small Poznan, Poland \\
\small pawel.kryszkiewicz@put.poznan.pl}
}

\maketitle

\begin{abstract}
In this paper, we consider the use of radio environment maps (REMs) in vehicular dynamic spectrum access (VDSA) for vehicle platooning applications. We propose an algorithm that dynamically allocates the frequency bands and transmission power in the so-called TV white spaces (TVWS) for intra-platoon messaging, intending to maximize the reliability of the communications, simultaneously keeping the interference to the primary system below the required threshold. The proposed solution is evaluated in simulations, with the results indicating a significant increase in communications reliability with VDSA. \footnote{Copyright © 2021 IEEE. Personal use us permitted. For any other purposes, permission must be obtained from the IEEE by emailing pubs-permissions@ieee.org. This is the author’s version of an article that has been published in the proceedings of 2021 IEEE 93rd Vehicular Technology Conference (VTC2021-Spring) and published by IEEE. Changes were made to this version by the publisher prior to publication, the final version of record is available at: http://dx.doi.org/10.1109/VTC2021-Spring51267.2021.9449077. To cite the paper use: P. Sroka, A. Kliks, M. Sybis and P. Kryszkiewicz, "Dynamic Power and Frequency Allocation Scheme for Autonomous Platooning," 2021 IEEE 93rd Vehicular Technology Conference (VTC2021-Spring), 2021, pp. 1-6, doi: 10.1109/VTC2021-Spring51267.2021.9449077  or visit https://ieeexplore.ieee.org/document/9449077}
\end{abstract}


\section{Introduction}
One of the key aspects of vehicles' communication is to ensure the safety of all road users. Among the many ways to do that is to use the platooning concept, where a~group of vehicles move in a~coordinated manner, like a~"road train", led by a~leader. In addition to improving safety, the use of a~platoon provides other gains: an increase in road capacity \cite{LPT+16}, reduction of fuel consumption  and, consequently, lower CO2 emissions~\cite{SARTRE}. In order to maximize these gains, vehicles should travel in very short distances at high speed, where automated mobility control is applied to ensure proper reaction time to changing conditions. An example of such a controller is the cooperative adaptive cruise control (CACC), where, apart from the data provided from on-board sensors, wireless communication is employed to exchange information between vehicles.

The communications between the vehicles can be achieved by means of dedicated short-range communications (DSRC). The concept of the DSRC system introduced in the United States as the WAVE standard (Wireless Access in Vehicular Environments) and in Europe as ITS-G5, uses the commonly known IEEE 802.11p standard \cite{IEEE80211}. The disadvantages of this solution, however, are its limited transmission range (typically up to about 300 m) and sensitivity to the hidden node phenomenon. Moreover, IEEE 802.11p is susceptible to the channel congestion problem when a high density of simultaneously transmitting vehicles is experienced.  Therefore, an extension of the Long Term Evolution (LTE) standard has been proposed for vehicle-to-vehicle (V2V) applications, known as the C-V2X (Cellular-V2X) \cite{VUKAD2018}. C-V2X is characterized by the greater range and reliability, but its full potential can only be used in the presence of cellular network infrastructure, and is also sensitive to the channel congestion problem \cite{VUKAD2018}.

The problem of channel blocking increases gradually with the new V2V applications introduced, such as the different safety systems or infotainment, as more and more data has to be distributed in the network. A~possible way to solve this issue is to use the vehicular dynamic spectrum access (VDSA) technique, which allows for a~dynamic change of frequency band used for V2V communications, including the use of licensed bands, such as the TV white spaces (TVWS)~\cite{SYB2018}. The use of VDSA allows to reduce the traffic in the 5.9~GHz band used by DSRC or C-V2X and to mitigate the impact of channel blocking. However, with the use of VDSA in the licensed bands one should account for the need for protection of the primary system performance. Thus, in this paper, we propose a VDSA algorithm with a power control mechanism applied to IEEE 802.11p transmitters, with the maximal allowable transmit power constrained in order to minimize the interference introduced to the primary system. Additionally, to account for the presence of additional interference from the licensed transmission, we dynamically modify the sensing detection threshold for the carrier-sense multiple access with collision avoidance (CSMA-CA) mechanism to minimize the chance of transmission blocking due to the presence of the primary system signal. To provide the required information on the licensed system we consider the use of radio environment maps (REMs), where the parameters of the primary user are stored in a~database with respect to the geographical location.

The paper is organized as follows. A system model is given in Section II. Section III outlines the problem formulation, while the dynamic resource allocation and the use of REMs is described in Section IV. Finally, the performance evaluation followed by conclusions are provided in Sections V and VI, respectively.

\section{System Model}
Within this investigation, we focus on an analysis of a~motorway scenario, where multiple platoons of cars, controlled using CACC algorithm \cite{Raj2000}, travel among other vehicles. Moreover, we consider the presence of fixed-location digital terrestrial television (DTT) receivers around the motorway, that require protection of their TV signal. We also assume that each platoon is preceded by a~jammer car that periodically changes its velocity from 130 km/h to 100 km/h and then back to 130 km/h, with a~single cycle duration of 30 s.
The autonomous platooning with CACC uses wireless communications between vehicles, where messages with mobility information either from the preceding car or from the platoon leader are transmitted. We assume that all cars communicate according to the IEEE 802.11p standard \cite{IEEE80211}, with the non-platoon cars broadcasting Cooperative Awareness Messages (CAMs) in the control channel (CCH) at 5.9 GHz, while the platoon vehicles, apart from CAMs transmitted in CCH, also send dedicated CACC messages. In addition, frequency bands, such as TVWS, are dynamically selected for intra-platoon communications. We assume that all platoon cars are equipped with dual radio, allowing for simultaneous operation in two frequency bands. For modeling of pathloss in the 5.9 GHz band we use the dual-slope model described in \cite{cheng2007}, while for the TVWS channels, in spite of lack of V2V channel models for these frequencies, we use the formula proposed in \cite{kkrrit}. \\
We employ VDSA technique to facilitate the transmission of CACC messages by the selection of additional frequency bands that the platoon will use as a~secondary system. The aim of the VDSA is to dynamically choose a~band that will allow maximizing the signal-to-interference-and-noise ratio (SINR) of the V2V transmission, simultaneously keeping the interference to the primary system at an acceptable level. To support the decision process of VDSA we assume it is aided with a~context information (e.g. on the observed TV signal power level) provided by REM databases.

\section{Problem description}
Following our prior works, i.e., \cite{Sroka2020a,Sroka2020b}, we define the joint power and frequency allocation problem. In particular, while assigning spectrum for a~platoon and while deriving the maximum acceptable transmit power, one has to guarantee successful data transmissions between platoon members, but also the incumbent users have to be protected. In our case, the primary users are mainly DTT receivers, which may be located close to the high-speed road. 
First, let us assume that there are $M$ DTT towers deployed in the considered area, which are broadcasting signals in $W$ TV channels. The bandwidth of the single TV channel depends on the country's regulations, for example, in Europe it is set to 8 MHz, whereas in the USA it is 6 MHz. Next, we consider the presence of $K$ platoons, and within each $k$-th platoon ($k=1,2,...,K$) there are $I_k$ transmitters and $J_k$ receivers. From the perspective of reliable communications within the entire platoon, we would like to maximize the minimum observable probability of detection within the platoon, i.e., we would like to improve maximally the worst link between the platoon leader and the worst-located platoon member. This approach is equivalent to the maximization of the worst signal to interference plus noise ratio (SINR) observed in the worst-located car. Please note that the worst-position within the platoon in terms of received-signal quality changes in time. The observed SINR at $j_k$-th receiver ($j_k=1,2,...,J_k$) while $i_k$-th transmitter ($i_k=1,2,...,I_k$) is sending data can be represented as 
\begin{equation}
SINR_{j_k,i_k}=\frac{P_{i_k}\left|h_{i_k,j_k}\right|^{2}}{N+I^{PU-V}_{j_k}+I^{V-V}_{j_k}}
\end{equation}
where $P_{i_k}$ is the transmit power of $i$-th transmitter in $k$-th platoon, $\left|h_{i_k,j_k}\right|^{2}$ is the wireless channel propagation gain (including antenna gains) of the $i_k$ - $j_k$ link. Next, $N$ and $I^{PU-V}_{j_k}$ is the thermal noise power and the interference power observed at $j_k$ receiver from the primary DTT users utilizing bands of interest, respectively.  $I^{V-V}_{j_k}$ is the interference power from all active (transmitting) cars in the $K$ platoons. Assuming that $f_k$ is the center frequency used by the $k$-th platoon, the interference component $I^{PU-V}_{j_k}$ depends on the adjacent channel interference ratio (ACIR) between the $m$-th DTT transmitter ($m=1,2,...,M$) using channel $w$, and the $k$-th platoon, denoted as $f^{ACIR,DTT-V}_{m,w, k}$. In our calculations we force that ACIR takes values within $\langle 0;1\rangle$ and is a~linear value describing the coupling between transmission and reception considering both emitted  signal spectrum (described by adjacent channel leakage ratio, ACLR, or spectrum emission mask, SEM) and the ability of the receiver to reject adjacent channel signal (described by adjacent channel selectivity, ACS). Focusing on occupied TV channels $w$ ($w=1,2,...,W$), and denoting the measured power of $m$-th DTT by $j_k$-th platoon member with $P^{DTT}_{m,w,j_k}$, the interference from primary system can be computed as \cite{Sroka2020b}: 
\begin{equation}
I^{DTT-V}_{j_k}=\sum_{m=1}^{M} \sum_{w=1}^W
P^{DTT}_{m,w,j_k} f^{ACIR,DTT-V}_{m,w,k},
\end{equation}
where $f^{ACIR,DTT-V}_{m,w,k}$ is a~function of the distance in frequency domain between the central frequencies of the $w$-th channel used by $m$-th DTT transmitter and $k$-th platoon. Assuming that all DTT transmitters use proprietary hardware compliant with all regulations in-force, the ACIR function may be approximated as independent from $m$.
Analogously, the interference from other active platoons' members, being a~random variable describing the functioning of the CSMA-CA algorithm, can be calculated as:
\begin{equation}
I^{V-V}_{j_k}=\sum_{l=1}^{K} \sum_{\substack{i_l \\ i_l \neq i_k}}
Pr\left(i_k\mid i_l\right)P_{i_l}
\left|h_{j_k,i_l}\right|^{2} f^{ACIR,VV}_{l,k}
\label{eq_int_v_v_basic}
\end{equation}
where $f^{ACIR,VV}_{l,k}$ is the ACIR value being the function of the frequency distance between the central frequencies allocated to platoons $l$ and $k$. $Pr\left(i_k\mid i_l\right)$ is the probability that $i_k$ transmits while $i_l$ is active. This conditional probability depends on the parameters of CSMA-CA protocol, e.g., the threshold for assessment of a~given channel as free or occupied.
The ultimate requirement for the secondary system, i.e., the platoons transmitting in TVWSs, is to protect the DTT receivers \cite{OFCOM_TVWS_2015, DSA_ruler_DTT_2017}. When the location of protected DTT receivers is not known, the worst case assumption should be made following e.g. \cite{DSA_ruler_DTT_2017}. In this case, DTT receiver distanced by 60 m from the transmitting vehicle can be considered. The DTT receiver has to be protected if the received DTT signal is usable, i.e., the power is above a~certain threshold $\Gamma$. Thus, the $w$-th DTT channel utilized by $m$-th DTT transmitter is considered as used by primary system in the neighborhood of vehicle $i_k$ if the observed power $P^{DTT}_{m,w,i_k}$ is above this threshold, i.e., $P^{DTT}_{m,w,i_k}>\Gamma$. Moreover, a~signal-to-interference ratio (SIR) at any DTT receiver utilizing channel $w$ has to be above certain threshold, i.e.,
\begin{align}
   & \forall{w,m,i_k:P^{DTT}_{m,w,i_k}>\Gamma}
    \nonumber \\ &
    P_{i_k}|h_{m,i_k}^{DTT}|^2 f^{ACIR,V-DTT}_{m,w,k}    <P^{DTT}_{m,w,i_k}/SIR_{\min}^{DTT},
\end{align}
where $|h_{m,i_k}^{DTT}|^2$ is channel gain between $i_k$ transmitter and a~potential $m$-th DTT RX, $f^{ACIR,V-DTT}_{m,w,k}$ is a~linear ACIR between platoon transmitter $i_k$ and $m$-th DTT reception at channel $w$. The value $SIR_{\min}^{DTT}$ is the minimum required SIR for DTT signal and platoon-based interference. 

In our scheme, we consider the possibility of application of the transmit power control mechanism by the platoon members. In particular, the transmit power is no longer fixed to some certain value but can be adjusted dynamically to the changing situation in the environment. We consider here that the transmit power may be changed in linear scale from 0 to some maximum value denoted as $P_{i_k}^{\max}$, i.e. $P_{i_k} \in \langle 0 ; P_{i_k}^{\max} \rangle$
Finally, the optimization problem can be defined as follows:
\begin{align}
  &  \max_{\substack{f_k \\ P_{i_k} }} \min_{j_k} SINR_{j_k,i_k}
\nonumber \\ &
s.t. \forall{m,w,i_k: P^{DTT}_{m,w,i_k}>\Gamma_{DTT}}
    \nonumber \\ &
    P_{i_k}|h_{i_k}^{DTT}|^2 f^{ACIR,V-DTT}_{m,w,k}
    <P^{DTT}_{m,w,i_k}/SIR_{\min}^{DTT},
    \nonumber \\ &
    P_{i_k} \in \langle 0 ; P_{i_k}^{\max} \rangle .
\end{align}
In the following sections, we provide the heuristic solution to this complicated optimization problem. 

\section{Dynamic resource allocation using REMs}
In the case of spectrum sharing, it is necessary to implement highly reliable spectrum sensing algorithms in order to protect the primary users (e.g. DTT receivers) and to detect the presence of other secondary transmissions. However, in the case of the platoons traveling along the high-speed roads, typically comprising at least two lanes in both directions, both the topographic surroundings and the radio environment may change very quickly due to the high travelling speed. Thus, it is beneficial to support the dynamic spectrum access procedures by exchanging supplemental data between the platoons and the infrastructure. In particular, the information about the DTT signal strength in a given location as well as historical statistics may be stored in the so-called radio environment map (REM). 
\subsection{Radio Environment Maps}
REM can be treated as the system of dedicated databases distributed over a certain area that stores information about the surrounding radio environment, such as signal strength or the location and the radio-characteristics of the radio transceivers. Besides being just the advanced set of repositories for radio data, REMs are typically considered to be equipped with some inferring capabilities, and some decision making engine, whose purpose is to support dynamic spectrum access process with conformance to the regulatory rules in force. The architecture of the REM-based supporting system may be treated as fully centralized (where one central entity processes all the delivered data and as such performs binding decisions), fully distributed (where each REM stores only information associated with local small vicinity), or a~hybrid one. 
REMs have been considered in various scenarios, mainly for spectrum sharing in TVWS \cite{Sodagari2015}, but also for e.g. wireless local-area network (WLAN) deployment \cite{Zou2020}, or recently for vehicular communications \cite{Katagiri2017,Sroka2020a}. Moreover, the concept of REMs (where typically the information only about the electromagnetic field was stored) has been extended to radio service maps (RSM) \cite{Tengkvist2017}, which may store various kinds of data, not related to radio environment but describing the generic situational context. 
In our research, we assume that the information about the distribution of the DTT reception power may be stored in the database jointly with other statistics like historical channel occupancy. As described in detail in \cite{Sroka2020a}, we have carried out intensive measurements of the received DTT signal while travelling with cars along the motorway. These data have been averaged over the number of conducted measurement campaigns, leading to the creation of the map of the DTT signal power.     

\subsection{Derivation of optimal sensing threshold}
Before starting transmission, each platoon member needs to decide on channel availability by performing carrier sensing. The channel may be treated as idle even if there is a~DTT signal present, thus  appropriate sensing threshold $\gamma$ shall be derived that considers the knowledge of DTT power \cite{Yucek2009,Tandra2008}).  At a certain location on the road, the observed signal can be defined as 
\begin{equation}
        y(t)=\left\{
                \begin{array}{ll}
                  h(t) * s(t) + n(t) + i(t) & \mathcal{H}_1\\
                  n(t) + i(t) & \mathcal{H}_0
                \end{array}
              \right. ,
\end{equation}
where $h(t) * s(t)$ represents the convolution of the signal $s(t)$ transmitted by any of the platoon transmitters $i_k$ with the channel impulse response $h(t)$. Next, $n(t)$ defines the additive white Gaussian noise (AWGN) with zero mean and variance $\sigma_n^2$, i.e. $n(t)\sim \mathcal{N}(0,\sigma_n^2)$. Finally, $i(t)$ represents the signal from the neighboring DTT tower, and we assume that from the perspective of the platoon receiver it will be treated as noise. In consequence, $i(t)\sim \mathcal{N}(\mu_i,\sigma_i^2)$. For the simplicity of notation, let us denote the observed power of the signal transmitted by other $i_k$ car as $\sigma_{s,k}^2$. Thus, the positive hypothesis $\mathcal{H}_1$ represents the presence of the transmission within platoons, and negative one $\mathcal{H}_0$ - the prospective vacancy of the channel (meaning lack of other transmission within platoons). Let's assume that the signal $y(t)$ is sampled to calculate the average received power, and $N_s$ samples are collected. Thus, the decision variable can be defined as $T(y) = \frac{1}{N_s}\sum_{i=0}^{N_s-1} |y(i)|^2$.
Thus, for both hypotheses, one can write that 
$T(y\mid \mathcal{H}_0) \sim \mathcal{N}\left(\sigma_i^2 + \sigma_n^2; \frac{2}{N_s}(\sigma_i^2 + \sigma_n^2)^2\right),
$ and $T(y\mid \mathcal{H}_1) \sim \mathcal{N}\left(\sigma_{s,k}^2+\sigma_i^2 + \sigma_n^2; \frac{2}{N_s}(\sigma_s^2 +\sigma_i^2 + \sigma_n^2)^2\right).
$
Fixing the decision threshold to some certain value $\gamma$, one may calculate the expected probability of false alarm $P_{fa}$ (i.e., a decision on the presence of the signal when the channel is vacant) and the probability of detection $P_d$ (i.e., correct detection of the present signal) as:
\begin{equation}
    P_{fa} = \Pr (T(y)>\gamma \mid \mathcal{H}_0) =  Q\left( \frac{\gamma - \sigma_i^2 - \sigma_n^2}{\sqrt{\frac{2}{N_s}}(\sigma_i^2 + \sigma_n^2)}\right)
\end{equation}
\begin{equation}
    P_{d} = \Pr (T(y)>\gamma \mid \mathcal{H}_1) =  Q\left( \frac{\gamma - \sigma_s^2 - \sigma_i^2 - \sigma_n^2}{\sqrt{\frac{2}{N_s}}(\sigma_{s,k}^2 + \sigma_i^2 + \sigma_n^2)}\right),
\end{equation}
where $Q(\dot)$ is the Marcum-Q function. Considering constant false alarm (CFAR) scheme, the detection threshold $\gamma$ equals: 
\begin{equation}
\label{eq:sensing1}
    \gamma = (\sigma_i^2 + \sigma_n^2)\left( \sqrt{\frac{2}{N_s}} Q^{-1}(P_{fa}) + 1 \right).
\end{equation}
Finally, when considering jointly the pair of probabilities $(P_{fa}, P_d)$, the necessary number of samples to be collected can be derived as $
    N_s = \frac{2}{SINR}\left(Q^{-1}(P_{fa}) - Q^{-1}(P_{d})(1+SINR)  \right)$.
In an ideal case the expected values of both probabilities may be achieved while observing the channel sufficiently long regardless of the current SINR value. It means, however, that for low SINR the sensing time may be too long for the vehicular scenario. Moreover, following findings from e.g. \cite{Tandra2008}, in real scenarios, the estimates of noise variance as well as the DTT power may be inaccurate. In consequence, the \textit{SNR wall} exists, it is impossible to achieve assumed pair of probabilities $(P_{fa},P_{d})$ if the observed SINR is too low. In order to deal with these problems, in the proposed VDSA algorithm we utilize the REMs, where the long-term measurements of the DTT power are stored. 
\subsection{Proposed VDSA Algorithm} 
In this work, we investigate the performance of intra-platoon communications where every vehicle is equipped with a~dual-radio transceiver, which allows for simultaneous operation in two different frequency bands. The first radio, denoted as \textit{Radio:0}, is constantly tuned to the 5.9 GHz control channel to be able to transmit and receive CAM messages. The second radio, denoted as \textit{Radio:1} is used only for the purpose of intra-platoon communication in TVWS using dedicated CACC messages, with the utilized frequency band selected dynamically using Algorithm \ref{algo}. The algorithm follows the reasoning presented in section III, maximizing the minimum SINR for all platoons. To obtain the information on the interference to and from the DTT system, REM databases are utilized. We assume for simplicity that the VDSA optimization is performed periodically every 1~s and jointly for all platoons in a~centralized way and the distribution of all required information (band selection notification, location reporting, etc.) between the spectrum management unit and the platoons is perfect. The proposed solution involves some degree of suboptimality, as the estimation of platoon-to-platoon interference described in~(\ref{eq_int_v_v_basic}) requires knowledge of the conditional probability that cars are transmitting at the same time instant, which is a nontrivial task. Therefore, as the probability of an event that more than two cars are transmitting simultaneously using IEEE 802.11p is very small, we approximate the value of $I_{i_{k}, f_k}^{V-V}$ by finding the worst-case (highest power) interference from the closest vehicle among other platoons.

\begin{algorithm}[!ht]
\caption{REM-based VDSA Algorithm}
\label{algo}
\begin{algorithmic}[1]
\Procedure {VDSA-Algorithm}{$K$, $M$, $W$ , $RX_{DTT}$,$F$,$\Gamma$,$SIR_{min}^{DTT}$ } \newline
\Comment {$K$ - set of platoons,$M$ - set of DTT transmitters, $W$ - set of DTT bands used by primary system, $RX_{DTT}$ - set of DTT receivers, $F$ - set of available frequency bands, $\Gamma$ - minimum protected DTT power threshold, $SIR_{min}^{DTT}$ - minimum SIR threshold for protected DTT.}
\For{each set of frequencies $(f_1, f_2, \ldots, f_N), f_n \in F, N = |K|$}:
\For{each platoon $k \in K$}:
\For {each DTT band $w \in W$ used by $m \in M$}:
\State Calculate $f_{m,w,k}^{ACIR,V-DTT}$ assuming $k$ uses freq. $f_{k}$.
\State Calculate $f_{m,w,k}^{ACIR,DTT-V}$ assuming $k$ uses freq. $f_{k}$.
\EndFor
\State Find the maximum transmit power constraint for the platoon leader vehicle $l_k$ as $P_{l_{k}, f_k}^{MAX} = \min_{w} (P_{w,l_{k}}^{MAX} / f^{ACIR,V-DTT}_{m,w,k})$, where  $P_{w,l_{k}}^{MAX} = \mathrm{\textsc{maxUEpower}}(l_{k}, w, RX_{DTT}, \Gamma, SIR_{min}^{DTT})$.

\For {each vehicle $i_{k}$ in platoon  $k$, $i_{k} \neq l_{k}$}:
\State Find the max. transmit power constraint as: $P_{i_{k}, f_k}^{MAX} = \min_{w} (P_{w,i_{k}}^{MAX} / f^{ACIR,V-DTT}_{m,w,k})$, where $P_{w,i_{k}}^{MAX} = \mathrm{\textsc{maxUEpower}}(i_{k}, w, RX_{DTT}, \Gamma, SIR_{min}^{DTT})$.
\State Calculate the interference from primary system: $I_{i_{k}, f_k}^{DTT-V} = \sum_{m=1}^{M}\sum_{w=1}^{W} P^{DTT}_{m,w,i_k} f^{ACIR,DTT-V}_{m,w,k}$.
\State Calculate the interference from other platoons: $I_{i_{k}, f_k}^{V-V} = \max_{i_p, p \in K, p \neq k} P_{i_p}^{MAX} |h_{i_p,i_k}|^2 f^{ACIR,V-V}_{p,k} $.
\State Find the minimum SINR of $i_k$ as: \newline
$SINR_{i_k, f_k}^{MIN} = \frac{\min(P_{l_{k}, f_k}^{MAX}|h_{l_k,i_k}|^2, P_{(i-1)_{k}, f_k}^{MAX}|h_{(i-1)_k,i_k}|^2)}{I_{i_{k}, f_k}^{DTT-V} + I_{i_{k}, f_k}^{V-V} + N}$

\EndFor
\State Find the minimum SINR in platoon $k$ as:\newline
$SINRp_{k, f_k}^{MIN} = \min_{i_k} SINR_{i_k, f_k}^{MIN}$
\EndFor
\EndFor
\State Find the best set of frequencies $(f_1, f_2, \ldots, f_N)^{*}$ as:\newline
$(f_1, f_2, \ldots, f_N)^{*} = \arg\max_{(f_1, f_2, \ldots, f_N)} \min_{k} SINRp_{k, f_k}^{MIN}$
\For{each platoon $k \in K$}:
\For {each vehicle $i_{k}$ in platoon  $k$}:
\State Configure \textit{Radio:1} with selected channel $f_k^{*}$.
\State Set the transmit power as: $P_{i_{k}, f_k^{*}}^{MAX}$.
\State Calculate $\gamma$ according to (\ref{eq:sensing1}).
\EndFor
\EndFor
\EndProcedure
\algstore{vdsa}
\end{algorithmic}
\end{algorithm}

\renewcommand{\thealgorithm}{}

\begin{algorithm}[!ht]
\caption*{Function \textsc{maxUEpower} used in Algorithm \ref{algo} }
\begin{algorithmic}[1]
\algrestore{vdsa}
\Procedure {maxUEpower}{$i_{k}$, $w$ , $RX_{DTT}$,  $\Gamma$, $SIR_{min}^{DTT}$}
\For {each DTT receiver $r$ in $RX_{DTT}$ }:
\State Find the received DTT signal power: $P_{m,r}^{DTT}$.
\State Calculate the estimated SIR for $r$ assuming $i_{k}$ is transmitting with full power:\newline
$SIR_{w,r}^{DTT} = \frac{P_{w,r}^{DTT}}{P_{i_{k}}^{MAX}|h_{w,r,i_k}^{DTT}|^2}$,
\If {$(SIR_{w,r}^{DTT} < SIR_{min}^{DTT}) \And (P_{w,r}^{DTT} > \Gamma)$}
\State $P_{w, r,i_{k}} \gets \frac{P_{i_{k}}^{MAX}SIR_{w,r}^{DTT}}{SIR_{min}^{DTT}} $
\Else
\State $P_{w, r, i_{k}} \gets P_{i_{k}}^{MAX}$
\EndIf
\EndFor
\State Find the maximum allowed transmit power for $i_{k}$ transmitting in band $w$ as:
$P_{w,i_{k}}^{MAX} \gets min_{r} (P_{w, r, i_{k}})$
\EndProcedure
\end{algorithmic}
\end{algorithm}

\section{Evaluation}
\subsection{Simulation scenario and parameters}
To evaluate the performance of VDSA in TVWS for platooning we conducted system-level simulations using a~simulation tool developed in C++ in the framework of analysis described in \cite{Sroka2020a, Sroka2020b}. We considered a~6-lane motorway, with two platoons moving on the outer lanes in opposite directions, with non-platoon cars distributed on the other lanes according to the uniform distribution with the average densities of 20 cars/km/lane or 50 cars/km/lane considered. Non-platoon cars transmitted CAMs every 100 ms in the CCH at 5.9 GHz, while the platoon vehicles, apart from the CAM messaging, could use dedicated CACC packets transmitted in TVWS for intra-platoon communications. Three use cases were considered:
\begin{enumerate}[label=(\alph*)]
    \item platoon cars transmit only CAMs at a~rate of 10 Hz.
    \item platoon cars transmit CAMs and CACC packets, both at a~rate of 5~Hz (aggregate messaging frequency is 10 Hz); the frequency band is selected according to strategy proposed in \cite{Sroka2020a}, without power control applied.
    \item as (b), but the TVWS band is selected according to Alg. \ref{algo} using power control and sensing threshold optimization.
\end{enumerate}
We assumed the existence of two DTT transmitters using center frequencies at 490 MHz and 522 MHz, respectively. The considered range of frequencies for VDSA spanned between 490~MHz  and  522~MHz with the resolution (frequency step) of 1~MHz.  The frequency selection was performed periodically every 1 s, with the duration of a~single simulation run set to 140 s. The considered threshold values for DTT signal protection were as follows: the minimum protected DTT receive power $\Gamma = -80$ dBm, and the minimum required SIR  level $SIR_{min}^{DTT} = 39.5$ dB \cite{DSA_ruler_DTT_2017}. We assumed that there are 10 DTT receivers present at specific locations around the motorway, with their coordinates and the observed DTT signal power levels stored in REM databases for use in VDSA. 

\subsection{Simulation results}
The main aim of applied VDSA for platooning is to improve the reliability of intra-platoon communications. Therefore, we investigated the reception ratio of transmitted packets between the members of the platoon, with the aim to obtain over 99\% successful receptions. Fig. \ref{fig_cam_rec} and \ref{fig_cacc_rec} present the observed successful reception ratio of platoon leader's packets vs. car position in platoon for CAM and CACC messages, respectively. We present only the results for the leader's packets, as the communications between the two consecutive cars (reception of preceding car packets) were much more reliable (reception rate above 99\%). From Fig. \ref{fig_cam_rec} one can conclude that with moderate traffic (20 cars/km/lane) it is almost possible to reach a high level of reliability using only the CCH at 5.9 GHz. However, when considering dense traffic (50 cars/km/lane) the performance in the 5.9 GHz band dropped significantly, with the observed successful reception rate even below 93\%. It improved when we applied offloading of some information to CACC messages transmitted in dynamically selected TVWS bands. Depending on the used strategy of selecting transmit power, we got 2\% gain in reception rate of CAMs when using full transmit power and 1\% when power control mechanism was applied. Moreover, with Fig. \ref{fig_cacc_rec} one can notice that the overall reception ratio improved significantly with VDSA applied, as the probability of successful reception of the leader's CACC packets is above 99.9\% with moderate traffic and above 99.8\% with dense traffic. Thus, we can conclude that offloading of the intra-platoon messages to the dynamically selected TVWS bands results in a significant increase in the successful reception ratio of leader's packets and can guarantee to reach the required reliability levels.\\
\begin{figure}[!b]
\centerline{\includegraphics[width=0.5\textwidth]{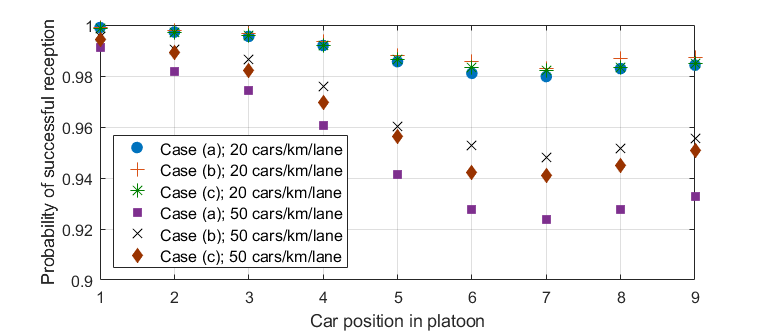}}
\caption{Reception rate of leader's CAM packets.}
\label{fig_cam_rec}
\end{figure}
\begin{figure}[!t]
\centerline{\includegraphics[width=0.5\textwidth]{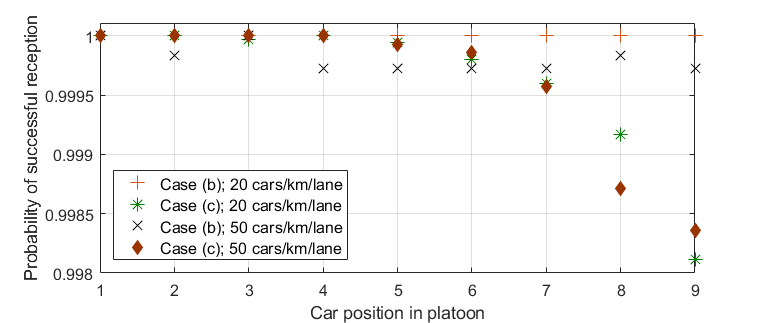}}
\caption{Reception rate of leader's CACC packets.}
\label{fig_cacc_rec}
\end{figure}
As the intra-platoon communication in the TVWS bands is a~secondary system, we have to ensure that the primary system operation is sufficiently protected. Therefore, we collected the SIR values observed in the simulation runs for the DTT signal reception at the selected locations of primary receivers, with the empirical cumulative distribution function (ECDF) of SIR shown in Fig. \ref{fig_sir}. With the analysis of the SIR ECDF one can conclude, that the use of TVWS for intra-platoon communications certainly can be harmful to the DTT transmission, with the observed SIR values for case~(b), where maximum transmit power is used, dropping well below the required threshold (60\% or 80\% of measurements below the required level, depending on the considered DTT band). However, when power control is applied according to the procedure described in Algorithm \ref{algo}, the SIR levels for almost all measurements can be kept above the threshold, with the several receptions being below the required SIR level due to the shadowing effect on channel gain estimation. Another gain observed with power control applied is that the average number of frequency band changes is much lower than in the case of using maximum power, with the values between 8-9 and 22-25 observed per simulation run with power control or with maximum power used, respectively.
\begin{figure}[!t]
\centerline{\includegraphics[width=0.5\textwidth]{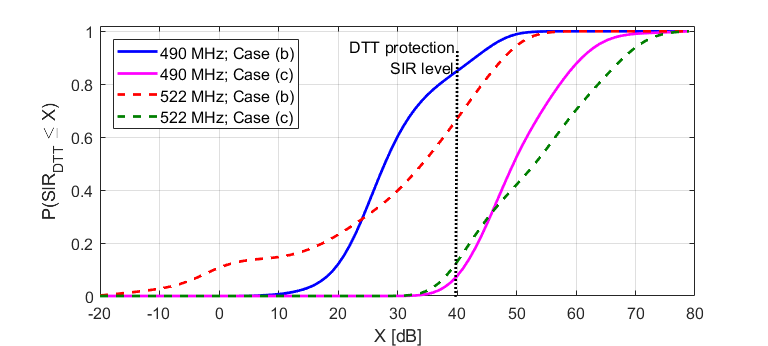}}
\caption{CDF of DTT receivers' SIR for the considered DVB-T bands.}
\label{fig_sir}
\end{figure}
Concluding the simulation results, the proposed solution to VDSA in TVWS using Algorithm \ref{algo} is able to improve the reliability of intra-platoon communications due to the capability of data traffic offloading to less occupied frequency bands. Moreover, with the power control mechanism applied it is possible to keep the interference to the primary DTT system below the required threshold. The gains of VDSA come at a~small cost of increased signalling, where additional context information on the DTT signal power levels and receivers' locations has to be provided from REMs.

\section{Conclusions}
The reliability of wireless communications is of the highest importance in the context of autonomous driving. We have shown that by offloading traffic from the congested 5.9 GHz band to the TVWS significant gain can be observed in terms of successful reception probability. As the transmission in TVWS is carried out on a spectrum sharing basis, where DTT receivers have to be protected from harmful interference, we have saved the characteristics of the DTT signal in the REM database, and utilized this information for power and frequency allocation in the proposed VDSA algorithm. Such an approach resulted in high reliability of data transmission within the platoon that increases the safety of autonomous mobility and, consequently, can result in a reduction of inter-vehicle distances. We foresee that in the future access to rich context information stored in REM may further improve the overall system performance and is subject to our further study.



\bibliographystyle{IEEEtran}
\bibliography{main.bib}

\end{document}